\documentclass[prb,twocolumn,amsmath,amssymb,floatfix]{revtex4}
\usepackage{graphicx}
\newcommand{\be}{\begin{eqnarray}}
\newcommand{\ee}{\end{eqnarray}}
\newcommand{\bfr}{{\bf r}}

\newcommand{\bfk}{{\bf k}}

\newcommand{\ha}{\hat{a}}

\newcommand{\hn}{\hat{n}}

\newcommand{\hatpsi}{\hat{\psi}}
\newcommand{\wbe}{\begin{widetext}}
\newcommand{\wee}{\end{widetext}}
\newcommand{\oncite}{\onlinecite}

\begin{document}
\draft

\title{Interaction-induced first order correlation between 
spatially-separated 1D dipolar fermions}

\author{Chi-Ming Chang$^{(1),(2)}$, Wei-Chao Shen$^{(2)}$, 
Chen-Yen Lai$^{(2)}$, 
Pochung Chen$^{(2),(3)}$ and Daw-Wei Wang$^{(2),(3)}$}

\address{
$^{(1)}$Physics Department, Harvard University, Cambridge, MA 02138
\\
$^{(2)}$Physics Department,
National Tsing-Hua University, Hsinchu, Taiwan 300, ROC
\\
$^{(3)}$Physics Institute, National Center for Theoretical Science,
Hsinchu, Taiwan 300, ROC
}

\date{\today}

\begin{abstract}
We calculate the ground state properties of fermionic dipolar
atoms/molecules
in a 1D double-tube potential by using the Luttinger liquid theory 
and the Density Matrix Renormalization Group calculation.
When the external field is applied near a magic angle with 
respect to the double-tube plane, the long-ranged 
dipolar interaction can generate a spontaneous correlation
between fermions in different tubes, 
even when the bare inter-tube tunneling rate is negligibly small.  
Such interaction-induced correlation strongly
enhances the contrast of the interference fringes, and therefore
can be easily observed in the standard time-of-flight experiment.
\end{abstract}


\maketitle
\section{Introduction}
Recent progress of ultracold atoms has made it possible to study
strongly correlated physics in a much wider parameter range.
One of the most important subjects is the one-dimensional (1D) physics, 
and many interesting phenomena, including Tonks-Girardeau gas
[\oncite{Bloch and the others}], Luttinger liquid (LL) behavior 
[\oncite{cazallila}], and polaronic effects in 
Bose-Fermi mixture [\oncite{mathey_wang}], etc. 
have been theoretically proposed or experimentally observed.
However, due to the short-range nature of
atomic interaction, it is not easy to study how the interaction 
between particles in different 1D tubes can bring different many-body effects.
In traditional solid state systems, on the other hand, the long-ranged
Coulomb interaction between electrons does induce several important
many-body features in low dimensional multi-component
systems: for example, Coulomb drag
between 1D double wires [\oncite{1d_drag_exp}], inter-wire/well 
coherence [\oncite{wang,sankar}], and spontaneous ferromagnetism (or exciton 
condensation) in double layer quantum Hall systems [\oncite{QH}]. 
Following these extensive studies in the the solid state systems, 
it is therefore very interesting to investigate how the long-ranged
dipolar interaction between ultracold atoms/molecules 
[\oncite{Cr,JILA_fermions}] can bring
emergent many-body physics in a spatially separated multi-component
system, where the first-order and second-order correlation function
can be observed easily in a time-of-flight (TOF) measurement. 
We note that several pioneering 
works have been reported to explore the exotic quantum phases 
of bosonic dipolar atoms/molecules [\oncite{wang_dipolar_liquid_Santos}]
in the multi-layer/tube systems, but results for fermionic dipoles seem
not well-explored yet to the best of our knowledge.

\begin{figure}
\includegraphics[width=8cm]{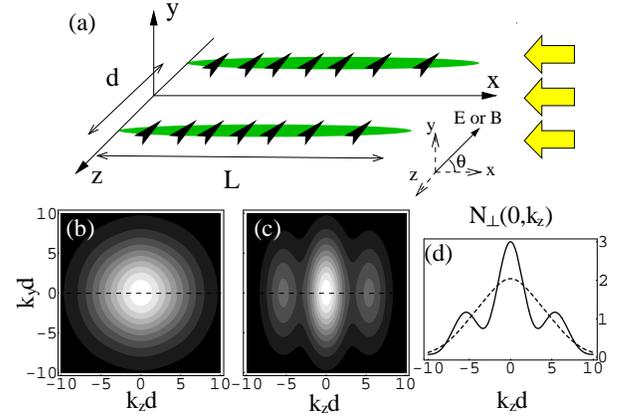}
\caption{
(Color online) (a) The double-tube system considered in the paper. Tilted arrows 
indicate the dipole moment of dipolar atoms/molecules. 
An external field (${\bf E}$ for the electric field or ${\bf B}$
for the magnetic field) is 
applied with an angle $\theta$ in the $x-y$ plane.
The big leftward arrows represent the imaging light
for taking a TOF image in the fast expanding (i.e. $y-z$) plane.
(b) and (c) are the calculated momentum distribution 
($N_\perp(k_y,k_z)$, proportional to the 
TOF image),  based on the DMRG result of the 
lattice model (see Eq. (\ref{H_lat}) and (\ref{N_perp}) below).
(b) is the noninteracting result, and (c) is the interacting result
with $\theta=\theta_c$. All the parameters used here are described 
in the text. The dashed and 
solid curves in (d) are the amplitude
profile along the horizontal line ($k_y=0$) of (b) and (c) respectively.
}
\label{system}
\end{figure}
In this paper, we investigate a 
1D double-tube system (see Fig. \ref{system}(a), and the similar setup for
bosonic atoms in Ref.[\oncite{interference}]), where 
{\it fermionic} polar molecules [\oncite{JILA_fermions}] (or magnetic 
dipolar atoms, $^{53}$Cr [\oncite{Cr}]) are loaded in the 
ground state of the confinement potential.
When no external field is applied, fermions are noninteracting and have
no correlation between fermions in the two tubes. Therefore
the time-of-flight (TOF) image (imaging light is 
along the tube direction) should be structureless (see Fig. \ref{system}(b)).
However, when an external electric/magnetic field is applied 
and tilted near a magic angle, $\theta_c=\cos^{-1}\sqrt{1/3}$, 
the intra-tube dipolar interaction is strongly
reduced, leaving only a repulsive and long-ranged 
{\it inter-tube} interaction. We show that such exotic 
interaction can induce the {\it first-order correlation}
between {\it fermions} in the two tubes, which
can be easily measured from the interference fringes
(see Fig. \ref{system}(c)). We note that unlike the 
interference between bosons 
[\oncite{interference,interference_boson}], the interference between such
spatially separated fermionic particles has no counterpart in classical waves, and is 
induced purely by the strongly correlated effect.

To investigate this problem systematically, in the following
we will first apply the celebrated LL theory with the 
renormalization group (RG) study in the weakly
interacting limit, and then use Density Matrix Renormalization
Group (DMRG) method in the strong 
interaction regime. The weak or strong interaction regime is defined
by the ratio between dipolar interaction to the kinetic energy.
Results from both calculations 
assure the existence of such interaction-induced inter-tube correlation,
and provide a rich quantum phase diagram.
For the convenience of later discussion, 
we will introduce the pseudo-spin up/down index for the 
upper/lower tube, so that the conventional definition of magnetic orders 
can be used to identify the quantum phases in the present system. 
For example, the ground state with spontaneous inter-tube correlation 
can be understood as an in-plane ferromagnetic order, 
$\langle\hat{\psi}^\dagger_{\uparrow}(x)
\hat{\psi}^{}_\downarrow(x)\rangle\neq 0$, where 
$\hat{\psi}_{\uparrow/\downarrow}(x)$ is the fermionic operator in
the upper/lower tube. As a result, the inter-tube correlation can 
be understood as a 1D planar ferromagnetic (FM) state with the 
pseudo-spin polarized in the $x-y$ plane of spin space. This is
exactly the 1D version of the well-known spontaneous ferromagnetism in 
double layer quantum Hall systems [\oncite{QH}]. Similar problem about
1D ferromagnetism of itinerant fermions is also an important subject
discussed in the literature [\oncite{1DFM}].
Note that the pseudo-spin defined here has 
nothing to do with the original spin of fermionic dipoles.

\section{Hamiltonian and Luttinger liquid theory in 
the weakly interacting limit}

Throughout this paper, we will assume that particles are loaded in 
the lowest subband of each tube with a transverse 
confinement wavefunction 
$\phi_s(y,z)=\frac{1}{\sqrt{\pi}R}e^{-(y^2+(z-sd)^2)/2R^2}$ 
and a Gaussian radius $R$ (here $s=\pm\frac{1}{2}
=\uparrow/\downarrow$ is the pseudo-spin index).
The resulting effective 1D system Hamiltonian then
can be written to be $H=H_0+H_I$, where $H_0$ is the kinetic energy:
\be
H_0=\sum_{s}\int_0^L dx\hatpsi^\dagger_s(x)
\frac{-\hbar^2}{2m}\partial_x^2\hatpsi_s^{}(x)
\ee
with $m$ being the mass of dipolar particles, and $H_I$
is the interaction energy: 
\be
H_I&=&\frac{1}{2}
\sum_{s,s'}\int_0^L dx\int_0^L dx'V_{|s-s'|}(x-x')
\nonumber\\
&&\times\hatpsi^\dagger_s(x)\hatpsi^{}_s(x)
\hatpsi^\dagger_{s'}(x')\hatpsi^{}_{s'}(x').
\ee
Here $V_{|s-s'|}(x)$ is the dipolar interaction 
between molecules in the same tubes ($V_0$) and/or in different 
tubes ($V_1$), obtained by integrating out the transverse
degree of freedom: 
$V_{|s-s'|}\left(x\right)\equiv
\int d{\bf r}_{1,\perp}\int d{\bf r}_{2,\perp}
\left|\phi_s(\bfr_{1,\perp})\right|^2
\left|\phi_{s'}(\bfr_{2,\perp})\right|^2
V_{d}\left({\bf r}_1-\bfr_2\right)$,
where $V_{d}(\bfr)=D^2(1-3(\hat{r}\cdot\hat{E})^2)/|\bfr|^3$ 
is the bare dipolar interaction
with $D$ being the electric dipole moment in c.g.s unit.
$\hat{E}$ is the unit vector along the external electric field.
Since in general the electric dipole 
interaction is much stronger (and tunable) than the 
magnetic dipole interaction, in the the rest of
this paper we will use polar molecules as the underlying particles 
for further discussion.
Extension to the magnetic dipolar atoms is straightforward.

Defining $\widetilde{V}_{0/1}(k)\equiv\int dx V_{0/1}(x)\,e^{-ikx}$ 
to be the Fourier transform of interaction, we can calculate 
their zero momentum ($k=0$) value analytically by integrating over 
the transverse confinement wavefunction, $\phi_s(\bfr_\perp)$:
\be
\widetilde{V}_{0}\left(0\right) &=&
\frac{D^{2}\left(1-3\cos^{2}\theta\right)}{R^2}
\label{V_0}
\\
\widetilde{V}_{1}\left(0\right)&=&D^{2}\left(
2-\left(2+\frac{d^{2}}{R^{2}}\right)e^{-\frac{d^{2}}{2R^{2}}}
\right)\frac{\sin^{2}\theta}{d^2}.
\label{V_1}
\ee
It is very easy to see that when $\theta\sim\theta_c=\cos^{-1}\sqrt{1/3}$
the intra-tube interaction is almost zero while the inter-tube interaction is
still finite and positive. Such interesting kind of interaction matrix element
cannot be realized in the traditional solid state system and therefore 
may bring some physics not predicted before. 

In the weakly interacting limit, we can apply the standard LL theory 
[\oncite{LL_review}] by linearizing the band energy 
around the two Fermi points, $\pm k_F$, and 
dividing fermions into the left/right movers (i.e. $\hatpsi_s(x)=
\sum_{r=\pm}\hatpsi_{r,s}(x)$, where $r=\pm$ is the chiral index).
The resulting low energy effective Hamiltonian can be rewritten to be, 
$H_{\rm eff}=H_{\rm LL}+H_1$, where
\be
H_{\rm LL}&=&v_F\int_0^L dx\sum_{r,s}\hatpsi^\dagger_{r,s}(x)
\left(-ir\hbar\frac{\partial}{\partial x}-k_{F}\right)\hatpsi_{r,s}(x)
\nonumber\\
&&+\frac{1}{2}\sum_{r,r',s,s'}\tilde{g}_{s,s'}^{r,r'}
\int_0^L dx\hat{\rho}_{r,s}(x)\hat{\rho}_{r',s'}(x)
\ee
is the LL Hamiltonian with the density operator,
$\hat{\rho}_{r,s}(x)\equiv\hatpsi^\dagger_{r,s}(x)
\hatpsi_{r,s}(x)$ and the forward scattering amplitude:
$\tilde{g}_{s,s'}^{r,r'}\equiv\widetilde{V}_{|s-s'|}(0)-g_{\|}
\delta_{r,-r'}\delta_{s,s'}$; 
\be
H_1\equiv g_{\bot}\sum_{s}
\int_0^L dx \hatpsi^\dagger_{+,s}(x)\hatpsi^{}_{-,-s}(x)
\hatpsi^\dagger_{-,-s}(x)\hatpsi^{}_{+,s}(x)
\ee
is the backward scattering between particles of different tubes and 
different chiralties. Here $v_F$ is the Fermi velocity, and 
$g_{\|/\perp}\equiv\tilde{V}_{0/1}(2k_F)$ is the backward scattering
amplitude. According to the standard Luttinger theory [\oncite{LL_review}],
the linearized band structure about the two Fermi points is
justified only when the interaction strength is much smaller than
the Fermi energy, i.e. when 
\be
\frac{D^2}{R^3}\ll \frac{\hbar^2 k_F^2}{2m}.
\label{condition_LL}
\ee

It is well-known that the LL Hamiltonian, $H_{\rm LL}$, 
can be diagonalized exactly via a Bogoliubov transformation, while
the backward scattering term, $H_1$, cannot be diagonalized in
the same basis. As a result, we have to use the standard 
one-loop renormalization group (RG) calculation [\oncite{LL_review}] 
to investigate when the later term is relevant
in the low energy limit, and how it renormalizes the former
one ($H_{\rm LL}$) in different parameter regime.
The bare Luttinger exponents are given by the initial system 
parameters, 
\be
K_{\rho/\sigma}\equiv\sqrt{
\frac{\pi v_{F}+\frac{1}{2}g_{\|}}
{\pi v_{F}+\widetilde{V}_{\rho/\sigma}-\frac{1}{2}
g_{\|}}},
\label{K_LL}
\ee
where $\widetilde{V}_{\rho/\sigma}\equiv
\widetilde{V}_0(0)\pm\widetilde{V}_1(0)$.
Within the one-loop RG, only $K_\sigma$ and the backward
scattering, $g_\perp$, are renormalized [\oncite{LL_review}],
following the RG equations below:
\be
\frac{dK_{\sigma}}{dl}=-\frac{1}{2}K^{2}_{\sigma}
\left(\frac{g_{\bot}}{\pi v_{\sigma}}\right)^{2},
\hspace{0.5cm}
\frac{dg_{\bot}}{dl}=2g_{\bot}
\left(1-K_{\sigma}\right),
\label{RG}
\ee
where 
\be
v_{\sigma}\equiv\sqrt{(v_{F}+\widetilde{V}_{\sigma}/\pi)^{2}
-(\widetilde{V}_{\sigma}/\pi)^{2}}
\label{v_spin}
\ee
is the collective mode velocity of 
the (pseudo-)spin-mode sector. $l=\ln\Lambda$ is the scaling parameter
with $\Lambda$ being the shortest length scale (or the largest momentum
scale) in the present system. For a given bare system parameter, 
$(K_\sigma,g_\perp)$, Eq. (\ref{RG}) then 
shows how they can flow to a fixed point, $(K_\sigma^\ast,
g_\perp^\ast)$, which determines the true low energy physical
properties of the double-tube system.

\section{Phase diagram in the Luttinger liquid theory}

To obtain a quantum phase diagram in the LL theory, we have to
use the RG result to calculate the scaling exponent, $\alpha$, of various 
correlation function in the long-distance limit, i.e. 
$\langle  \widehat{O}^\dagger(x)
\widehat{O}(0)\rangle\sim x^{-2+\alpha}$ as $|x|\to\infty$, where
$\widehat{O}(x)$ is the order parameter operator of interest.
For a given system parameter, the dominant order parameter is 
then determined by the most slowly decaying correlation function 
(or the largest and positive $\alpha$), known as the quasi-long-ranged order
in 1D systems. 
\begin{figure}
\includegraphics[width=8cm]{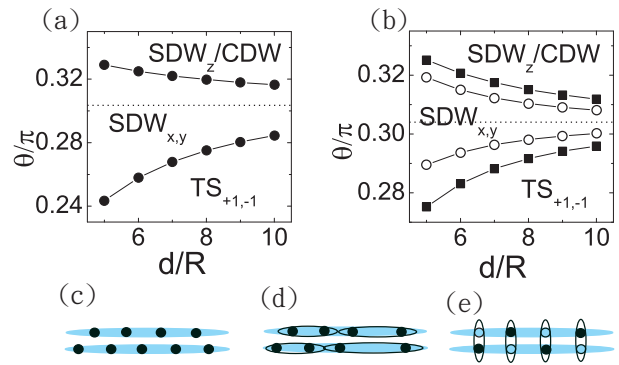}
\caption{(Color online) Quantum phase diagrams in terms of the tube separation
relative to the tube radius, $d/R$, and the tilted angle, $\theta/\pi$, in the LL theory. (a)$k_{F}R=0.1$ (filled circle), (b)$k_{F}R=0.2$ (filled square) 
and $k_{F}R=0.4$ (open circle).
The dashed line indicates the magic angle, $\theta_c=\cos^{-1}\sqrt{1/3}
\approx 0.304\pi$. 
(c)-(e) are cartoons to represent the axial SDW, 
polarized triplet superfluid, and the planar SDW phases respectively 
as defined in the text (also consistent with the literature 
[\oncite{LL_review}]). The filled and open circles here indicate 
the particle and hole fluctuation respectively. 
The the horizontal ellipses in (d) indicate
pairing between fermionic particles, while the
vertical ellipses in (e) indicates an inter-tube
correlation (i.e. there is an uncertainty for the particle position
between the two tubes). The average particle density distribution is
uniform for SDW$_{x,y}$ phase (e), while it is staggered for
SDW$_z$ phase (c), which has no inter-tube correlation.
} 
\label{phase_LL}
\end{figure}
Similar to the standard Luttinger liquid theory for other 1D systems 
[\oncite{LL_review}], we investigate the following candidates of order 
parameters [\oncite{LL_review}], because scaling exponents for other more
complicated kinds of order parameters are always smaller than
the ones below and therefore becomes negligible.
The definitions of order parameters we considered in this paper includes
(the associated scaling exponent is shown
after the definition):
Wigner crystal ($\widehat{O}_{\rm WC}=
\sum_s\hatpsi^\dagger_{+,s}\hatpsi^\dagger_{+,-s}\hatpsi^{}_{-,-s}
\hatpsi^{}_{-,s}$, $\alpha_{\rm WC}=2-4K_\rho$),
charge density wave ($\widehat{O}_{\rm CDW}=
\sum_s\hatpsi^\dagger_{+,s}\hatpsi^{}_{-,s}$, 
$\alpha_{\rm CDW}=2-K_\rho-K_\sigma$),
axial (pseudo-)spin density wave ($\widehat{O}_{{\rm SDW}_z}
\equiv\sum_{s,s'}\hat{\sigma}^z_{s,s'}
\hatpsi^\dagger_{+,s}\hatpsi^{}_{-,s'}$, $\alpha_{{\rm SDW}_z}=2-K_\rho-K_\sigma$),
planar (pseudo-)spin density wave ($\widehat{O}_{{\rm SDW}_{x,y}}
\equiv\sum_{s,s'}\hat{\sigma}^{x,y}_{s,s'}
\hatpsi^\dagger_{+,s}\hatpsi^{}_{-,s'}$, $\alpha_{{\rm SDW}_{x,y}}=2-K_\rho-
K_\sigma^{-1}$, and here $\hat{\sigma}^{x,y,z}_{ij}$ is 
Pauli matrix element), 
singlet superfluid ($\widehat{O}_{SS}
\equiv\sum_s s\hatpsi^{}_{+,s}\hatpsi^{}_{-,-s}$,
$\alpha_{SS}=2-K_\rho^{-1}-K_\sigma$), 
unpolarized triplet superfluid ($\widehat{O}_{TS_0}
\equiv\sum_s\hatpsi_{+,s}\hatpsi^{}_{-,-s}$, 
$\alpha_{TS_0}=2-K_\rho^{-1}-K_\sigma$),
and polarized triplet superfluid ($\widehat{O}_{TS_{2s}}
\equiv\hatpsi_{+,s}\hatpsi^{}_{-,s}$ for $s=\pm 1/2$,
$\alpha_{TS_{2s}}=2-K_\rho^{-1}-K_\sigma^{-1}$).
In the low energy limit, the exponent $\alpha$ for each phase is then obtained
from the renormalized Luttinger exponents, $K_\sigma^\ast$
(obtained via solving Eq. (\ref{RG})) and $K_\rho^\ast=K_\rho$.
Note that $K_\rho^\ast$ is not renormalized within the one-loop 
RG scheme and therefore is the same as the
initial(bare) value obtained from the system parameters.

After calculating the scaling exponents of various
order parameters mentioned above,
we find that only three of them can be the dominant phases in the 
system we consider here. They are axial (pseudo-)spin density wave
(SDW$_z$), planar (pseudo-)spin density wave (SDW$_{x,y}$),
and polarized triplet superfluid (TS$_{\pm}$). The charge density wave
(CDW) has the same scaling exponent as SDW$_z$, but it can be shown
that it still decay fast than the later case by a logarithmic correction
after integrating along the RG flow [\oncite{LL_review}]. In this paper,
we will still consider both of them to be the dominant phase, but only
discuss the physics related to the SDW$_z$ order for simplicity.
From the definition of scaling exponent above, it is easy to see that
the phase boundary between the planar (pseudo-)spin density wave (SDW$_{x,y}$),
and the polarized triplet superfluid (TS$_{\pm}$) is determined by 
$K_\rho^\ast=K_\rho=1$, while the phase boundary between 
the axial (pseudo-)spin density wave
(SDW$_z$) and the planar (pseudo-)spin density wave (SDW$_{x,y}$)
is by $K_\sigma^\ast=1$. From the expression of the bare LL exponent
in Eq. (\ref{K_LL}), we can find the first phase boundary is then 
determined by the following simple equation:
\be
\tilde{V}_\rho=\tilde{V}_0+\tilde{V}_1=g_\|=\tilde{V}_0(2k_F).
\label{boundary1}
\ee

However, when considering the phase boundary between SDW$_z$
and SDW$_{x,y}$, we have to calculate the renormalized Luttinger
exponent, $K_\sigma^\ast$, from the RG equation in Eq. (\ref{RG}).
Integrating Eq. (\ref{RG}) directly, it is easy to show that the solution
connecting to $K_\sigma^\ast=1$ and $g_\perp^\ast=0$ 
(required as a fixed point of Eq. (\ref{RG}), 
$dK_\sigma/dl=dg_\perp/dl=0$) 
is given by
\be
1-\frac{1}{K_\sigma}-\ln K_\sigma
=-\frac{1}{8}\left(\frac{g_\perp}{\pi v_\sigma}\right)^2.
\label{RG_line}
\ee
To obtain the functional dependence of the interaction strength on this
RG line, we can insert the bare form of $K_\sigma$ in Eq. (\ref{K_LL})
and take the weak interaction limit. The leading order terms then give
the following condition for the phase boundary between SDW$_z$
and SDW$_{x,y}$:
\be
\tilde{V}_\sigma&=&\tilde{V}_0-\tilde{V}_1
\nonumber\\
&=&g_\perp+g_\|=\tilde{V}_0(2k_F)+\tilde{V}_1(2k_F).
\label{boundary2}
\ee
Since each term in Eqs. (\ref{boundary1}) and (\ref{boundary2}) 
is proportional to $D^2$, the phase boundary lines determined above
are independent of the interaction strength (more precisely, true only 
in the weak interaction limit). As a result, the phase diagram of LL theory 
can be completely determined by the following three dimensionless 
parameters: $d/R$, $\theta$, and $k_FR$.

In Figs. \ref{phase_LL}(a) and (b), we show the calculated phase
diagram as a function of $d/R$ and $\theta$, for three different values
of particle density, $k_FR$.
For a given inter-tube separation, $d$, the system is dominated 
by the axial (pseudo-)spin density wave (SDW$_z$) in the regime
of large tilted angle $\theta$. Such state
can be also understood as a  ``phase-locked'' state with staggered
modulation of the particle density (see Fig. \ref{phase_LL}(c)), similar
to the 1D electron gas in the semi-conductor double wire system
[\oncite{stern}]. This phase results from the fact that 
when $\theta$ is large, the intra-tube
interaction is repulsive and stronger than the inter-tube interaction, 
making a crystalized density distribution in each tube.
At the same time, the backward scattering, $g_\perp$, 
scales to be infinite (i.e. $g_\perp^\ast\to\infty$ because 
$K_\sigma^\ast<1$, see Eq. (\ref{RG})), and therefore 
opens a single particle excitation gap. Such divergence of $g_\perp$
also leads to a vanishing $K_\sigma^\ast$ according to Eq. (\ref{RG}).
This is why the scaling exponent of SDW$_z$ becomes much larger
than SDW$_{x,y}$ in the low energy limit (see above for the definition of
scaling exponents for each phase).
In the regime of small $\theta$, on the other hand, the intra-tube 
interaction becomes attractive, leading to
a triplet superfluid ($TS_{\pm}$) (Fig. \ref{phase_LL}(d)). 
It is easy to see from the order parameters
of the above two phases, there is no inter-tube correlation, and 
therefore no visible interference fringes in the fast expanding plane, 
similar to the noninteracting result in Fig. \ref{system}(b).

However, when $\theta$ is in the intermediate range and 
close to $\theta_c\sim 0.304\pi$, the ground state 
becomes the planar (pseudo-)spin density wave (SDW$_{x,y}$).
In this parameter regime, the intra-tube interaction becomes
smaller than the inter-tube interaction (i.e. $|\tilde{V}_0(0)|<
\tilde{V}_1(0)$, see Eqs. (\ref{V_0})-(\ref{V_1})), 
and therefore the backward scattering
becomes {\it irrelevant} in the low energy limit ($g_\perp^\ast\to 0$
because $K_\sigma^\ast >1$, see Eq. (\ref{RG})), 
leading to a {\it uniform} density distribution along the tube. 
Besides, the order parameter,  $\widehat{O}_{{\rm SDW}_{x,y}}$, 
has implied that at each position on the tube, fermionic particles has equal 
and non-zero probability to be found in the upper and
the lower tube (or say, there is an uncertainty for a particle to be
in one tube or the other, although the average probability in these
two tubes are the same), showing an interaction induced inter-tube 
correlation (with a periodic modulation of the relative gauge phase,
see Fig. \ref{phase_LL}(e)). Although within the LL theory,
such inter-tube correlation is a quasi-long-ranged order, we still 
expect to observe it in the interference pattern (similar to
Fig. \ref{system}(c)) in a finite size system after integrating along the
tube direction, if only the scaling exponent, $\alpha$, is not too small. 
For typical parameters of polar molecules, say OH molecules 
($m\sim 17$ a.m.u. and the largest dipole moment is $D\sim 1.68$ Debye), 
we can consider $R\sim 0.1$ $\mu$m, $d\sim 0.5$ $\mu$m, and
$k_F\sim \pi\times 10$ $\mu$m$^{-1}$, and hance the obtained 
$\alpha_{SDW_{x,y}}$ can be
of the order of 0.1 or more, showing that the probability to observe
such inter-tube correlation is not small. We note that
this planar (pseudo-)SDW$_{x,y}$ phase with a
spontaneous inter-tube correlation is a completely new phase,
and has no counterpart even in the traditional solid state 
(semi-conductor or metal) double-wire system. 

\section{Ferromagnetism in strongly interacting regime}

In the strongly interacting regime (i.e. the interaction energy is 
of the same order of or even larger than Fermi energy), 
LL theory fails to give correct
low energy physics and hence we apply DMRG method to 
numerically study the ground state properties. 
For the convenience of numerical calculation, 
from now on we will consider $\theta=\theta_c$ only to eliminate the 
intra-tube interaction completely. Besides,
we assume an optical lattice is applied along the tube direction ($x$)
so that the system Hamiltonian can be written in 
a single band lattice model:
\be
H_{\rm lat}&=&-t\sum_{j,s}
\left(\ha^\dagger_{j,s}\ha^{}_{j+1,s}
+h.c.\right)-J\sum_{j,s}\ha^\dagger_{j,s}\ha^{}_{j,-s}
\nonumber\\
&&+U\sum_{j}\hn_{j,\uparrow}\hn_{j,\downarrow}
+\frac{V}{2}\sum_{\langle j_1,j_2\rangle,s}\hn_{j_1,s}\hn_{j_2,-s}
\label{H_lat}
\ee
where $\ha_{j,s}$ and 
$\hn_{j,s}=\ha^\dagger_{j,s}\ha^{}_{j,s}$ here are
the fermionic field operator and density operator of the site $j$ and
the pseudo-spin $s=\uparrow/\downarrow$. Here
$t$, $J$, $U$, and $V$ are the inter-site tunneling in the
same tube, the inter-tube tunneling, the inter-tube onsite 
repulsion, and the inter-tube nearest-neighboring-site repulsion
respectively. Note that we introduce inter-tube tunneling $J$ here just
for the completeness of numerical calculation, but we will just concentrate
on the physics in the limit of zero $J$ after the calculation, i.e.
in the limit of zero inter-tube tunneling.
The numerical values of these parameters can be easily calculated by
integrating over the on-site Wannier function, and can be 
tuned in a wide range by changing the external field strength, 
lattice strength, and/or lattice spacing. Since it is not our
purpose in this paper to provide a numerical comparison with any 
certain experimental setup (there is unfortunately no such system available
yet), we will not calculate their absolute
values in this paper, but will present all of our results in dimensionless
parameters instead. 
For simplicity, we also have neglected the interaction of longer 
(next-nearest-neighboring sites) range.

\begin{figure}
\includegraphics[width=8.cm]{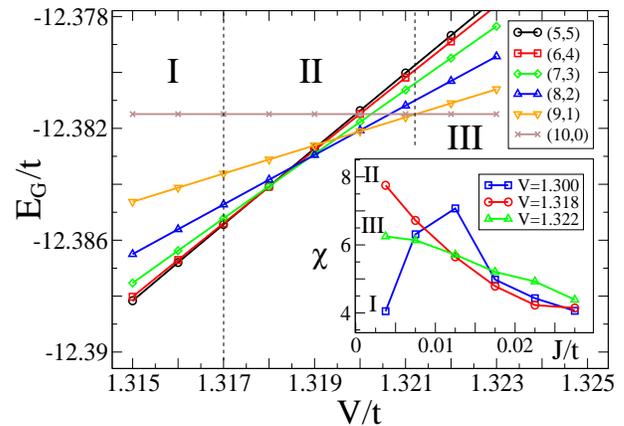}
\caption{(Color online) Ground state energy of Eq. (\ref{H_lat}) (for $J=0$) 
as a function of $V/t$ for different particle numbers in the two 
tubes, ($N_\uparrow,N_\downarrow$). 
Regime I, II, and III are the SDw$_{x,y}$, canted ferromagnetic
(or inter-tube correlated) phase, and axial ferromagnetic phase 
respectively. The detailed definition is in the text.
Inset shows the 
magnetic susceptibility $\chi$ (defined in the text) as a function 
of inter-tube hopping, $J$. 
}
\label{E_FM}
\end{figure}

In Fig. \ref{E_FM} we first show the ground state energy, $E_G$, 
as a function of $V$, by taking $J=0$, $U/t=3$ and the total
particles, $N=10$, in a 1D lattice of length $L=20$. Note that
by taking $J=0$, particle numbers in each tube is a conserved
quantity and therefore we can label the system eigenstates by
($N_\uparrow,N_\downarrow$).
There are three different regimes of interest: in regime I ($V<1.317$), 
$E_G$ increases linearly as a function of $V$, and the numbers of particles
in the two tubes are the same ($N_\uparrow=N_\downarrow=5$) in the
ground state. Detailed analysis shows that the ground state 
waevfunction is the same as
the SDW$_{x,y}$ phase in the weak interaction limit discussed above.
In regime III ($V>1.321$), $E_G$ becomes independent of $V$, because 
all particles are moved to either one of the tubes to minimize 
the strong inter-tube interaction energy. This phase can be described
as an axial ferromagnetic state by breaking the $Z_2$ (spatial inversion) 
symmetry of the Hamiltonian, Eq. (\ref{H_lat}). 
Inside the narrow window of regime II  ($1.317<V<1.321$), the 
ground state energy changes step by step as the particle number 
difference changes, indicating a series of first order phase transitions.
In the thermodynamic limit ($N,L\gg 1$), it is reasonable 
to expect that the ground state energy should become
a smooth curve in regime II, i.e. an {\it infinitesmall} 
inter-tube tunneling ($J\to 0$)
can easily mix many degenerate states and lead to a 
coherent ground state,
$|\Psi_C\rangle$, which has finite single particle inter-tube 
correlation between fermions in the two spatially separated tubes. 
Using the language of magnetism, we can define the pseudo-spin 
magnetization density to be
$\vec{\cal M}\equiv\frac{1}{N}\sum_j
\sum_{s,s'}\vec{\sigma}_{s,s'}\ha^\dagger_{j,s}\ha^{}_{j,s'}$,
and such spontaneous inter-tube correlation can be understood as an
in-plane ferro-magnetization, i.e. 
$\langle\Psi_C|{\cal M}_x|\Psi_C\rangle\neq 0$.

To confirm this conjecture, we further investigate results with
a finite but small inter-tube tunneling, $J$, which is equivalent to an
effective "magnetic field" along the $x$ direction in the 
{\it pseudo-spin space}. In other words, we can study
how the system magnetization responds to such small
perturbation by calculating the "magnetic" susceptibility, $\chi\equiv 
\partial \langle\Psi_C|{\cal M}_x|\Psi_C\rangle/
\partial J$. As shown in the inset of Fig.
\ref{E_FM}, $\chi$ has a tendency to diverge
as $J\to 0$ when $V$ is tuned to be inside the regime II, strongly indicating
an in-plane FM order. Combining the fact that the average numbers
of particles in the two tubes are also different in this regime (i.e.
$\langle\Psi_C|{\cal M}_z|\Psi_C\rangle\neq 0$, as indicated from
Fig. \ref{E_FM}), 
the ground state in regime II is best understood as a pseudo-spin
canted ferromagnetism, i.e. pseudo-spin is uniformly polarized with
a tilted angle about the  pseudo-spin $z$ axis. 
This is another striking new phase predicted 
in this paper, breaking a continuous $U(1)$ symmetry in 1D system. 
Different from the planar SDW discussed in the weakly interacting limit,
the pseudo-spin FM state obtained here in the strongly
interacting regime has a uniform (rather than periodic oscillation) 
magnetization. From the experimental point of view, this
order parameter indicates a strong interference fringes as shown
in Fig. \ref{system}(c).
Unfortunately, so far we have not been able to calculate systems
of larger size with an efficient code, and therefore the demonstration 
of the existence of such symmetry breaking state in the 
thermodynamic limit may need more justification.
However, to the best of our knowledge, our system is probably
the first {\it physical} candidate (rather than a mathematical 
model) to realize the ferromagnetism of 1D itinerant fermions:
it is not be restricted by the celebrated Lieb-Mattis theorem [\oncite{lieb}] 
or Mermin-Wigner theorem [\oncite{mermin}] due to the nontrivial 
long-ranged and anisotropic dipolar interaction. Numerical results 
for larger system size and more complete phase diagram of 
this double-tube system should be an interesting direction to explore
in the future.

\section{Interference pattern in the time-of-flight experiment}

Since the theme of this paper is to demonstrate
the existence of an interaction-induced inter-tube correlation between
1D dipolar fermions (true for both weak and strong interaction regime), 
the best experimental evidence is from 
the interference pattern between these fermions.
Theoretically, we can calculate the momentum
distribution function of the ground state and then integrate 
along the tube direction
to obtain the TOF image in the fast expanding ($y-z$) plane,
$N_\perp(\bfk_\perp)$ (here $\bfk_\perp\equiv(k_y,k_z)$, see Fig. 
\ref{system}). Since particles can fly fast from the tube center 
after the strong confinement potential is released, such TOF
image should eventually become equivalent to momentum distribution
after a long time of flight. After integrating out the contribution
along the tube direction ($x$), we can obtain the following
expression for $N_\perp(\bfk_\perp)$:
\be
N_\perp(\bfk_\perp)
&=&4\pi R^2 N e^{-|\bfk_\perp|^2R^2}
\left(1+2\langle{\cal M}_x\rangle\cos(k_z d)\right),
\label{N_perp}
\ee
where we have used a
Gaussian type-wavefunction (with radius $R$) to approximate 
the initial confinement wavefunction, $\phi_s(\bfr_\perp)$, and 
$\langle{\cal M}_x\rangle\equiv
\langle\Psi_C|{\cal M}_x|\Psi_C\rangle$ is the inter-tube correlation. 
When the inter-tube tunneling, $J$, is small and the interaction is weak, 
the calculated inter-tube 
correlation ($\langle{\cal M}_x\rangle$) is so small that the 
obtained momentum distribution (i.e. TOF image) is close to a 
broad Gaussian-like function without any structure 
(see Fig. \ref{system}(b) and the dashed line in 
Fig. \ref{system}(d)). On the other hand, when the direction and 
the strength of dipolar interaction is tuned to the regime II of 
Fig. \ref{E_FM}, the long-ranged inter-tube interaction strongly
enhances the inter-tube correlation ($\langle{\cal M}_x\rangle$) 
and hence the contrast of interference fringes. In 
Fig. \ref{system}(c) (and the solid line in Fig. \ref{system}(d)),
we show a typical calculated interference pattern (i.e. $N_\perp(\bfk)$)
with parameters in this regime: $U/t=3$, $V/t=1.318$, 
and $R/d=0.167$ for quarter-filling ($L=20$ and $N=10$). 
We find that even the same single particle inter-tube 
tunneling rate ($J/t=0.1$) is used in the calculation of 
both Fig. \ref{system}(b) and (c),
the obtained interference patterns are totally different: 
the later has strong contrast in the interference fringes 
due to the interaction effect.
Unlike the usual interference fringes between bosonic particles,
the fermionic dipoles here do not have a condensate 
in each tube, and therefore has
no counterpart in any classical waves. 

As for the quasi-long-ranged order predicted by the Luttinger liquid
theory in the weakly interacting regime, one should measure the TOF
image in the $x-z$ plane by using the imaging light perpendicular to
the double tube plane  ($y$). As a result both SDW$_z$ and SDW$_{x,y}$
phases can show some kinds of Bragg peaks along the $x$-axis
due to the periodic distribution of density and/or phase fluctuations, 
while the TS$_{\pm 1}$ phase does not have such feature. Furthermore,
SDW$_z$ phase has a single particle gap, while SDW$_{x,y}$ and
TS$_{\pm 1}$ do not have such gap. This gap can be easily measured
by light scattering spectroscopy and hence becomes 
a way to distinguish these
two density wave phases. We note that although the inter-tube
correlation of SDW$_{x,y}$ phase is periodically oscillating and the
correlation function decays as a power-law in large distance, the finite
tube length can still make it possible to measure some residual inter-tube
correlation, which does not exist in the other two phases. According
to the results of bosonic atoms [\oncite{interference,Mathey}], the noise
correlation of the TOF image for different lengths of tube can also
provide the experimental value of the Luttinger exponent.

\section{Summary}

In this paper, we find an interaction-induced
inter-tube correlation between spatially-separated fermionic polar 
molecules (or dipolar atoms) in the 1D double-tube potential.
Such correlation does not exist if the interaction between fermions
is too weak, showing a clearly many-body effect in a 1D system.
We use both Luttinger liquid theory (with proper renormalization group
method) to study the phase diagram in weak interaction limit, and
use Density Matrix Renormalization Group Method to calculate
the order parameters in strong interaction regime.
This new phenomenon predicted in this paper can be observed 
in the first order interference pattern, and has no counterpart either
in the classical waves or in the 1D solid state systems.

\section{Achknowledgement}

We thank S. Das Sarma, G. Fiete, I. Spielman, and H.-H. Lin for 
fruitful discussions. This work is supported by NSC via NCTS in Taiwan.



\begin{thebibliography}{99}

\bibitem{Bloch and the others}
B. Paredes, A. Widera, V. Murg, O. Mandel, S. F$\ddot{o}$lling, I. Cirac, G. V. Shlyapnikov, T. W. H$\ddot{a}$nsch and I. Bloch, Nature {\bf 429}, 277 (2004);
T. Kinoshita, T. Wenger, and D. S. Weiss, Science {\bf 305}, 1125 (2004).

\bibitem{cazallila}
For a recent review of Luttinger liquid theory in 1D Bose gas, see
M. A. Cazalilla, J. Phys. B: AMOP {\bf 37}, S1 (2004);
M. A. Cazalilla, A. F. Ho, and T. Giamarchi,
New J. Phys. {\bf 8}, 158 (2006).

\bibitem{mathey_wang}
L. Mathey, D.-W. Wang, W. Hofstetter, M. D. Lukin, and E. Demler, Phys. Rev. Lett. {\bf 93}, 120404 (2004);
L. Mathey and D.-W. Wang, Phys. Rev. A {\bf 75}, 013612 (2007).

\bibitem{1d_drag_exp}
P. Debray, V. Zverev, O. Raichev, R. Klesse, P. Vasilopoulos and R. S. Newrock, J. Phys. Condens. Matter {\bf 13}, 3389 (2001);
P. Debray, V. N. Zverev, V. Gurevich, R. Klesse and R. S. Newrock, Semicond. Sci. Technol. {\bf 17}, R21 (2002);
M. Yamamoto, M. Stopa, Y. Tokura, Y. Hirayama, S. Tarucha, Physica {\bf 12E} 726 (2002).

\bibitem{wang}
D.-W. Wang, E. G. Mishchenko, and E. Demler, Phys. Rev. Lett. 
{\bf 95}, 086802 (2005).

\bibitem{sankar}
L. Zheng, M. W. Ortalano, and S. Das Sarma,
Phys. Rev. B {\bf 55}, 4506 (1997);

\bibitem{QH}
I. B. Spielman, J. P. Eisenstein, L. N. Pfeiffer, and K. W. West,
Phys. Rev. Lett. {\bf 87}, 036803 (2001);
M. Kellogg, J. P. Eisenstein, L. N. Pfeiffer, and K. W. West, Phys. Rev. Lett. 
{\bf 90}, 246801 (2003); M. Kellogg, J. P. Eisenstein, L. N. Pfeiffer, and K. W. West, Phys. Rev. Lett. {\bf 93}, 036801 (2003).
For a review of bilayer QH effect, see 
S. M. Girvin and A. H. MacDonald, in {\it Perspectives in Quantum Hall Effects}
edited by S. Das Sarma and A. Pinczuk
(John Wiley \& Sons, New York, 1997); J. P. Eisenstein, {\it ibid}, and
reference therein.

\bibitem{Cr}
J. Stuhler1, A. Griesmaier, T. Koch1, M. Fattori, T. Pfau, S. Giovanazzi, P. Pedri, and L. Santos, Phys. Rev. Lett. {\bf 95}, 150406 (2005);
T. Lahaye, T. Koch, B. Fr$\ddot{o}$hlich, M. Fattori, J. Metz, A. Griesmaier, S. Giovanazzi, and T. Pfau, Nature {\bf 448}, 672 (2007); 
T. Koch, T. Lahaye, J. Metz, B. Fr$\ddot{o}$hlich, A. Griesmaier, and T. Pfau, Nature Physics {\bf 4}, 218 (2008). 

\bibitem{JILA_fermions}
K.-K. Ni, S. Ospelkaus, M. H. G. de Miranda, A. Pe'er, B. Neyenhuis, J. J. Zirbel, S. Kotochigova, P. S. Julienne, D. S. Jin, and J. Ye, Science {\bf 322}, 231 (2008).

\bibitem{wang_dipolar_liquid_Santos}
D.-W. Wang, M. D. Lukin, and E. Demler, Phys. Rev. Lett. {\bf 97}, 180413 (2006);
D.-W. Wang, Phys. Rev. Lett. {\bf 98}, 060403 (2007);
A. Arguelles and L. Santos, Phys. Rev. A {\bf 75}, 053613 (2007);
C. Kollath, Julia S. Meyer, and T. Giamarchi1, Phys. Rev. Lett. {\bf 100}, 130403 (2008).

\bibitem{interference}
A. Polkovnikov, E. Altman, and E. Demler, PNAS {\bf 103}, 6125 (2006);
V. Gritsev, E. Altman, E. Demler and A. Polkovnikov, Nature Phys. {\bf 2}, 705 (2006);
S. Hofferberth, I. Lesanovsky, T. Schumm, A. Imambekov, V. Gritsev, E. Demler, and J. Schmiedmayer, Nature Phys. {\bf 4}, 489 (2008).

\bibitem{interference_boson}
M. R. Andrews, C. G. Townsend, H.-J. Miesner, D. S. Durfee, D. M. Kurn, and W. Ketterle, Science {\bf 275}, 637 (1997).

\bibitem{1DFM}
E. Lieb, Phys. Rev. Lett. 62, 1201 (1989); A. Mielke,
J. Phys. A 24, L73 (1991); M. Ulmke, Eur. Phys. J. B 1,
301 (1998); T. Okabe, cond-mat/9707032; L. Bartosch, M. Kollar, and P. Kopietz, Phys. Rev. B 67, 092403 (2003); 
A. Mielke, Phys. Lett. A 174, 443 (1993); H. Tasaki, Phys. Rev. Lett. 75, 4678 (1995); S. Daul and
R. M. Noack, Phys. Rev. B 58, 2635 (1998), and reference
therein.

\bibitem{LL_review}
J. Solyom, Adv. Phys. {\bf 28}, 201 (1979); 
J. Voit, Rep. Prog. Phys. {\bf 58}, 977 (1995);
T. Giamarchi, {\it Quantum Physics in One Dimension}, 
(Oxford University Press, USA, 2004).

\bibitem{stern}
R. Klesse and A. Stern, Phys. Rev. B {\bf 62}, 16912 (2000);
V. V. Ponomarenko and D. V. Averin, Phys. Rev. Lett. {\bf 85}, 
4928 (2000).

\bibitem{lieb}
E. Lieb and D. Mattis, Phys. Rev. {\bf 125}, 164 (1962).

\bibitem{mermin}
N. D. Mermin and H. Wagner, Phys. Rev. Lett. {\bf 17}, 1133 (1966).

\bibitem{Mathey}
 L. Mathey, E. Altman, and A. Vishwanath,  Phys. Rev. Lett. 
 {\bf 100}, 240401 (2008).

\end{thebibliography}
\end{document}